\pdfoutput=1
\documentclass[aps,prl,floatfix,twocolumn,superscriptaddress]{revtex4-1}
\usepackage{physics}
\usepackage{amsfonts}
\usepackage{mathrsfs}
\usepackage{amsmath}
\usepackage{color}
\usepackage{graphicx}
\usepackage{bm}
\usepackage{amssymb}
\usepackage{xspace}
\usepackage{epstopdf}
\usepackage{dcolumn}
\usepackage{longtable}
\usepackage[colorlinks=true, letterpaper=true, pdfstartview=FitV, linkcolor=blue, citecolor=blue, urlcolor=blue]{hyperref}
\begin{document}
\title{High-Temperature Majorana Corner States}
\author{Qiyue Wang}
\affiliation{Department of Physics, University of Texas at Dallas, Richardson, Texas 75080, USA}
\author{Cheng-Cheng Liu}
\affiliation{School of Physics, Beijing Institute of Technology, Beijing 100081, China}
\author{Yuan-Ming Lu}
\affiliation{Department of Physics, Ohio State University, Columbus, Ohio 43210, USA}
\author{Fan Zhang}
\affiliation{Department of Physics, University of Texas at Dallas, Richardson, Texas 75080, USA}
\begin{abstract}
Majorana bound states often occur at the end of 1D topological superconductor
or at the $\pi$ Josephson junction mediated by a helical edge state.
Validated by a new bulk invariant and an intuitive edge argument,
we show the emergence of one Majorana Kramers pair at each corner
of a square-shaped 2D topological insulator proximitized by
an $s_\pm$-wave ({\em e.g.}, Fe-based) superconductor.
We obtain a phase diagram that emphasizes the roles of bulk parameters and edge orientations.
We propose several experimental realizations in lattice-matched candidate materials.
Our scheme offers a high-temperature platform for exploring higher-order non-Abelian quasiparticles.
\end{abstract}
\maketitle

\indent\textcolor{blue}{\em Introduction.}---A central theme in condensed matter physics is to
discover and classify distinctive states of matter.
Conventionally, states such as magnets or superconductors are characterized by
the time-reversal or gauge symmetry that they spontaneously break.
Over the last decade, the discovery of topological insulators (TI)
has opened the door to various classes of topological states of matter~\cite{Moore,Kane,Qi,Chiu}.
In each class, all the states respect the same symmetries, yet they are indexed by
the different values of a bulk topological invariant, which determine the physics
on their boundaries of one lower dimension. As a prime example,
for a 2D/3D TI, the nontrivial $\mathbb{Z}_2$ index of the insulting bulk state dictates
the presence of gapless 1D/2D edge/surface state.
When coupled to a magnet or superconductor that breaks an essential symmetry,
the boundary state may acquire an energy gap and even be passivated~\cite{QAH1,QAH2,TSC1,TSC2}.

Recently, a novel class of TIs coined ``higher-order TIs''~\cite{corner1,corner2,PW,WA,JL,ZS,FS,LL,MZ,MZ1} 
has emerged. They host protected gapless states on boundaries of more than one dimensions lower. 
For instance, a second-order 2D/3D TI has gapless corner/hinge states between distinct edges/surfaces that are gapped.
While the emergent corner states have been realized in a phononic quadrupole TI~\cite{corner2},
the prototype hinge states have been responsible for a quantum anomalous Hall effect~\cite{QAH2}.
These examples have enlightened the search for fascinating higher-order topological matter.

Meanwhile, a priority in condensed matter physics is to create topological superconductors (TSC)
with Majorana bound states~\cite{Read,Kitaev,Ivanov,Nayak},
which offer a decoherence-free platform for quantum computation and information.
A promising route~\cite{K,S,A,L,O,J,B,T,F,C} is to employ an architecture that proximity couples an ordinary superconductor
with a material that effectively has one helical band,
which requires a magnetic field or a $\pi$ Josephson junction.
While several experiments are achieving this goal~\cite{ex1,ex2,ex3,ex4,ex5},
one might wonder whether there exists a higher-order TSC~\cite{EK,MG}, 
whose nontrivial bulk topology leads to the emergence of Majoranas,
and how such a TSC can be realized in an experimentally accessible setup.

\begin{figure}[!t]
\includegraphics[width=0.90\columnwidth]{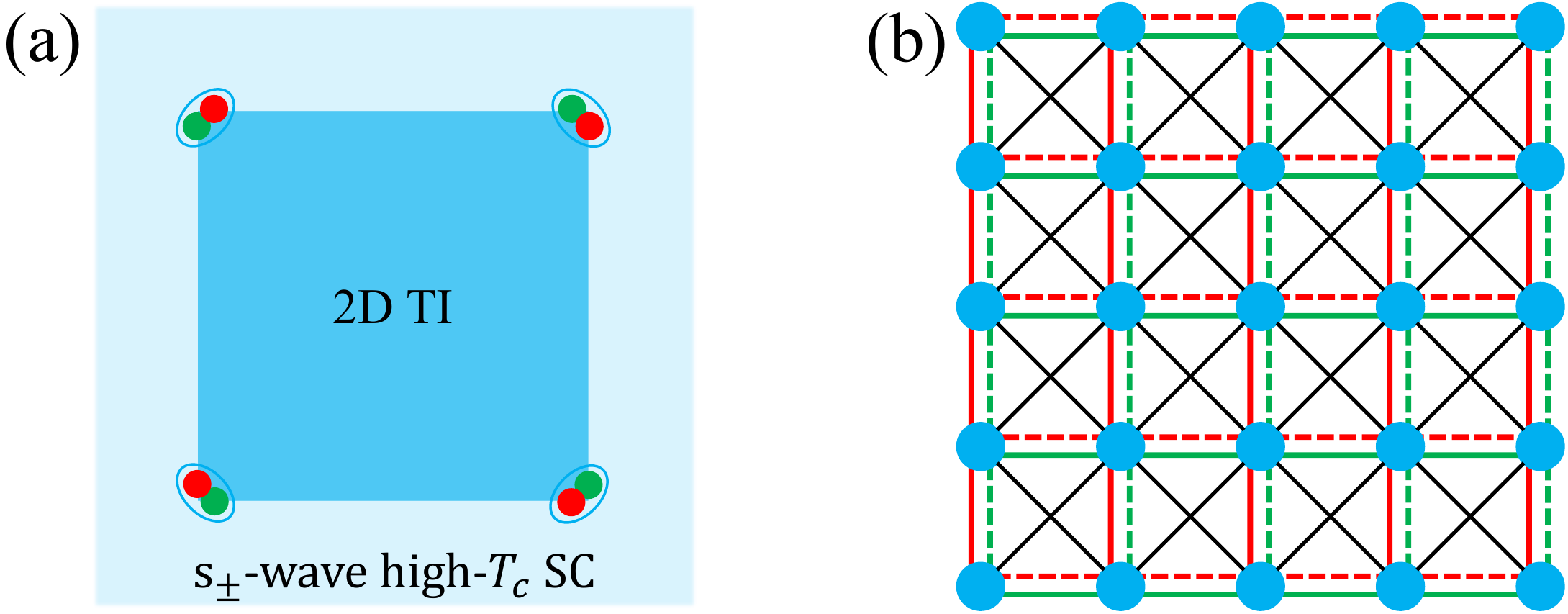}
\caption{Schematics of (a) a 2D TI proximitized by a nodeless high-$T_c$ 
({\em e.g.}, Fe-based $s_{\pm}$-wave) superconductor yielding corner Majoranas and
(b) the lattice model described by Eq.~(\ref{Hlattice}).}
\label{fig1}
\end{figure}

Here we show that a second-order TSC in class DIII ({\em i.e.}, time-reversal-invariant)
can be realized by proximitizing a 2D TI~\cite{Liu,Wrasse,Wan} with an $s_\pm$-wave
superconductor~\cite{Mazin,Bozovic,Dai}, as sketched in Fig.~\ref{fig1}a. 
While the bulk has an insulating gap and the edges acquire superconducting gaps,
there is a Majorana Kramers pair~\cite{Qi2,FZ1,Wong,Nakosai,Berg,Flen,Loss,Haim,Schrade}
at each corner. To demonstrate this TSC,
we not only provide an intuitive edge argument but also derive a novel bulk invariant
based on an emergent $\mathcal{C}_4$ symmetry. Moreover,
we obtain a general phase diagram and propose several experimental realizations.
Remarkably, our scheme requires neither a $\pi$ Josephson junction nor a magnetic field,
and the superconductor used here is topologically trivial and has a high critical temperature.
Our work establishes an unprecedented high-temperature platform
for exploring higher-order TSCs and Majoranas.

\indent\textcolor{blue}{\em Minimal model.}---We first introduce a time-reversal-invariant (TRI)
model with two orbitals per site in a square lattice, as sketched in Fig.~\ref{fig1}b,
to describe a 2D TI proximitized by an $s_\pm$-wave superconductor,
\begin{equation}
\begin{aligned}
H=&\Bigg(t \sum_{\langle ij\rangle_{x},s}-\;\;t \sum_{\langle ij\rangle_{y},s}
+\;\;t_1 \sum_{\langle\langle ij\rangle\rangle,s}\Bigg)c^{\dagger}_{i \mu s}\sigma_z^{\mu\nu}c_{j \nu s}\\
&+{i\lambda_{\rm R}}\sum_{<ij>}c^{\dagger}_{i\mu\alpha}\left(\bm{s}^{\alpha\beta}\times
\hat{\bm{d}}^{\,ij}\right)_z \sigma_x^{\mu\nu} c_{j\nu\beta}\\
&+\Delta_0\sum_{i,\sigma}c^{\dagger}_{i\sigma\uparrow}c^{\dagger}_{i\sigma\downarrow}
+\Delta_1\sum_{\langle ij\rangle,\sigma}c^{\dagger}_{i\sigma\uparrow}c^{\dagger}_{j\sigma\downarrow}
+\mbox{H.c.}.
\end{aligned}
\label{Hlattice}
\end{equation}
Here $\bm s$ and $\bm \sigma$ are the Pauli matrices for the spin and orbital spaces, respectively.
The $t$-term is the nearest-neighbor intra-orbital hopping, with opposite signs in the $\hat x$ and $\hat y$ directions.
The $t_1$-term is the next-nearest-neighbor intra-orbital hopping.
Both terms have opposite signs for different orbitals.
The $\lambda_R$-term arises from the inter-orbital Rashba spin-orbit coupling;
$\bm{d}^{\,ij}$ is a vector pointing from site $j$ to site $i$.
$\Delta_0$ and $\Delta_1$ combine to provide an $s_{\pm}$-wave pairing.

\begin{table}[b!]
\centering
\caption{\label{Table}
{Band inversion at the TRI momenta, TI $\mathbb{Z}_2$ index of $h_{\bm k}^{\rm TI}$,
and the second-order TSC $\mathbb{Z}_2$ index of $\mathcal{H}_{\bm k}^{\rm BdG}$
for the cases with $t>0$ and $\mu=0$. The cases with $t<0$ and $\mu=0$
can be obtained by switching the indices at $(\pi,0)$ and $(0,\pi)$.
$\pm$ denote the parity of the number of band inversions.}}
\newcommand\T{\rule{0pt}{2.5ex}}
\newcommand\B{\rule[-1.7ex]{0pt}{0pt}}
\centering
\begin{tabular}{ccccccc}
\hline\hline
Condition\quad & $\;(0,0)\;$ & $\;(\pi,0)\;$ & $\;(0,\pi)\;$ & $\;(\pi,\pi)\;$ &
$\;\mathbb{Z}_2$-TI\quad & $\;\mathbb{Z}_2$-TSC \T\\[3pt]
\hline
$t>t_1>0$ & $+$ & $-$ & $+$ & $+$ & $1$ & $1$ \T\\
$t_1>t>0$ & $+$ & $-$ & $-$ & $+$ & $0$ & $0$ \T\\
$t>-t_1>0$ & $-$ & $-$ & $+$ & $-$ & $1$ & $1$ \T\\
$-t_1>t>0$ & $-$ & $+$ & $+$ & $-$ & $0$ & $0$ \T\\
\hline\hline
\end{tabular}
\end{table}

It is more convenient to rewrite Eq.~(\ref{Hlattice}) as the following Bogoliubov-de Gennes (BdG) Hamiltonian
\begin{eqnarray}
\begin{aligned}
\mathcal{H}^{\rm BdG}_{\bm k}=&\left(h^{\rm TI}_{\bm k}-\mu\right)\tau_z+\Delta_{\bm k} \tau_x,\\
h^{\rm TI}_{\bm k}=&\;[2t(\cos k_x-\cos k_y)+4t_1 \cos k_x \cos k_y]\sigma_z\\
&+2\lambda_{\rm R} (\sin k_x s_y-\sin k_y s_x)\sigma_x,\\
\Delta_{\bm k}=&\;\Delta_0+2\Delta_1 (\cos k_x+\cos k_y),
\label{Hbdg}
\end{aligned}
\end{eqnarray}
where $\mu$ is the chemical potential and $\bm \tau$ are the Pauli matrices in Nambu particle-hole notation.
$\Delta_{\bm k}$ is the $s_{\pm}$-wave pairing that switches signs between
the zone center $\Gamma$ $(0,0)$ and the zone corner M $(\pi,\pi)$ when $|\Delta_0|<4|\Delta_1|$.
The 2D material can acquire such an $s_{\pm}$-wave pair potential, {\em e.g.},
when it is proximity coupled to a nodeless Fe-based high-temperature superconductor~\cite{Mazin,Bozovic,Dai}.
Importantly, our model Eq.~(\ref{Hbdg}) has time-reversal ($\Theta=is_y\mathcal{K}$), particle-hole
($\Xi=s_y\tau_y\mathcal{K}$), and inversion ($\mathcal{P}=\sigma_z$) symmetries,
with $\mathcal{K}$ being the complex conjugation. These symmetries can be expressed as follows
\begin{eqnarray}
\begin{gathered}
\Theta \mathcal{H}^{\rm BdG}({\bm k}) \Theta^{-1}=\mathcal{H}^{\rm BdG}(-{\bm k}),\\
\Xi \mathcal{H}^{\rm BdG}({\bm k}) \Xi^{-1}=-\mathcal{H}^{\rm BdG}(-{\bm k}),\\
\mathcal{P} \mathcal{H}^{\rm BdG}({\bm k}) \mathcal{P}^{-1}=\mathcal{H}^{\rm BdG}(-{\bm k}).\\
\end{gathered}
\label{Symm}
\end{eqnarray}

The 2D material is described by $h^{\rm TI}$ in Eq.~(\ref{Hbdg}) and respects
the time-reversal and inversion symmetries in Eq.~(\ref{Symm}).
The spectrum of $h^{\rm TI}$ is generally gapped except when $t_1=0$ or $|t|=|t_1|$.
Thus, the Fu-Kane criterion~\cite{P-Z2} based on the $\mathcal{P}$ eigenvalues at the four TRI momenta
can be used to evaluate whether the 2D material is a TI or not.
Since $h^{\rm TI}$ is the same at $(0,0)$ and $(\pi,\pi)$, the $\mathbb{Z}_2$ index is determined by
the relative band inversion from $(\pi,0)$ to $(\pi,0)$.
As listed in Table~\ref{Table}, the material is a $\mathbb{Z}_2$ TI for $|t|>|t_1|$
and a trivial insulator otherwise.

\begin{figure}[!t]
\includegraphics[width=1\columnwidth]{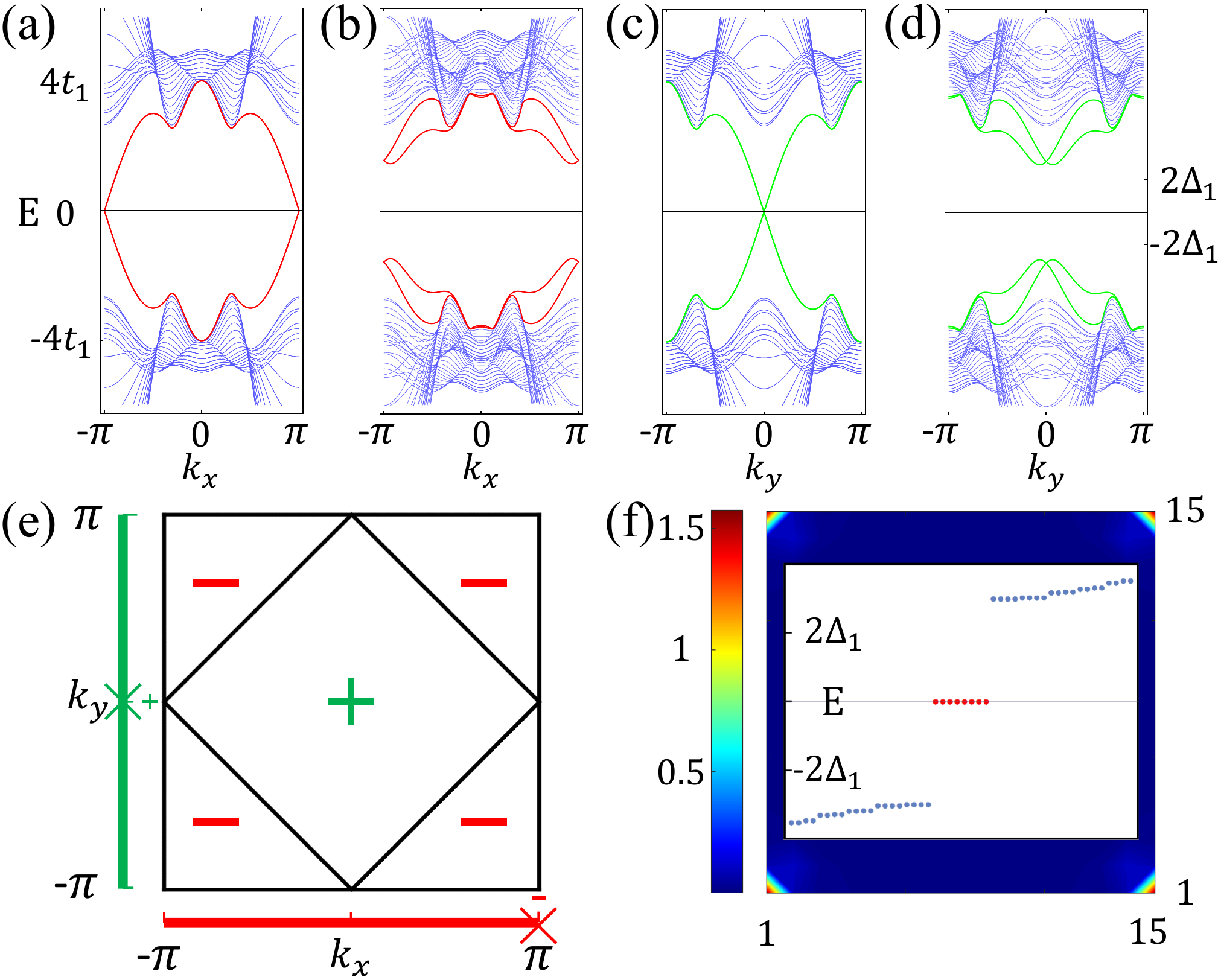}
\caption{(a) Band structure of a 2D TI ribbon exhibiting a $(0{\bar 1})$ helical edge state at $k_x=\pi$.
(b) BdG spectrum of (a) with an $s_{\pm}$-wave pairing.
(c)-(d) Similar to (a)-(b), but exhibiting a $({\bar 1}0)$ helical edge state at $k_y=0$.
(e) Schematic of the $s_{\pm}$-wave pairing in the bulk BZ
and the $s$-wave pairing of opposite signs acquired by the two edge states.
(f) Exact diagonalization revealing the four pairs of Majoranas for the $15\times15$ square size TI:
the density plot displays their corner localized probability distribution and the inset features their symmetry enforced zero energy.
We have chosen $t_1=1$, $t=2$, $\lambda_R=1.5$, $\Delta_0=0$, $\Delta_1=0.5$, and $\mu=0.5$ in all panels.}
\label{fig2}
\end{figure}

Consider the case in which the 2D TI has only one band inversion at $(\pi,0)$.
Figs.~\ref{fig2}a~and~\ref{fig2}c plot the band structures of TI ribbons
along the $\hat x$ and $\hat y$ directions.
As anticipated, the $(0{\bar 1})$ edge state emerges at $k_x=\pi$,
whereas the $({\bar 1}0)$ edge state emerges at $k_y=0$.
When the Fermi energy lies in the bulk gap,
the proximity induced $s_{\pm}$-wave pairing $\Delta_{\bm k}$ can gap out
both edge states by breaking the gauge symmetry,
as shown in Figs.~\ref{fig2}b~and~\ref{fig2}d for the $\Delta_0=0$ case. 
For such a proximitized TI in a square shape as sketched in Fig.~\ref{fig1}a,
in spite of the fact that the 2D bulk and the 1D edges are all fully gapped,
there are four zero-energy Majorana Kramers pairs --- one at each corner,
as exhibited by Fig.~\ref{fig2}f. Evidently, this realizes a second-order TSC in class DIII.

An edge argument can explain the presence of a local Majorana Kramers pair when $\mu$ is small,
as illustrated in Fig.~\ref{fig2}e. At the $(0{\bar 1})$ edge, the helical edge state at $k_x=\pi$
acquires a negative pairing since $\Delta_{\bm k}$ is negative at $k_x=\pi$ for all $k_y$'s.
By contrast, at the $({\bar 1}0)$ edge, the helical edge state at $k_y=0$
acquires a positive pairing since $\Delta_{\bm k}$ is positive at $k_y=0$ for all $k_x$'s.
In light of the topological criterion for 1D TSCs in class DIII~\cite{Qi2,FZ1},
such a pairing sign reversal leads to the emergence of a boundary Majorana Kramers pair.
This argument equally applies to the four corners in Fig.~\ref{fig2}f.

\indent\textcolor{blue}{\em Topological invariant.}---The edge argument is intuitive
in understanding the second order TSC,
yet the edge-state theory is only valid near the Dirac points.
(In fact, a helical edge state cannot be captured by any 1D lattice model.)
However, not only can $\mu$ be far from the Dirac points,
but $\Delta_{\bm k}$ also has a strong ${\bm k}$-dependence across the Brillouin zone (BZ).
Thus, it is necessary to establish a topological invariant by using the 2D bulk state.

The bulk model Eq.~(\ref{Hbdg}) is invariant under the spinful four-fold rotation, accompanied by a gauge transformation
that flips the signs of $\Delta_1$ pairing, 
odd-parity orbital on one set of $\sqrt{2}\times\sqrt{2}$ sublattices,
and even-parity orbital on the other set.
This composite symmetry reads
\begin{eqnarray}\begin{gathered}
\mathcal{C}_4\mathcal{H}^{\rm BdG}(k_x,k_y)\mathcal{C}_4^{-1}=\mathcal{H}^{\rm BdG}(\pi-k_y,\pi+k_x)
\label{C4}
\end{gathered}\end{eqnarray}
with $\mathcal{C}_4=\sigma_z\tau_z e^{-i s_z\pi/4}$ for $\Delta_0=0$.
Essentially, the four symmetry operators fulfill the following algebra
\begin{eqnarray}
\begin{gathered}
\Xi^2=\mathcal{P}^2=1,\quad\Theta^2=(\mathcal{C}_4)^{4}=-1,\\
[\Theta,\Xi]=[\Xi,\mathcal{P}]=[\mathcal{P},\Theta]=0,\\
[\mathcal{C}_4,\Theta]=[\mathcal{C}_4,\mathcal{P}]=0,\quad\{\mathcal{C}_4,\Xi\}=0.
\label{Comm}
\end{gathered}
\end{eqnarray}
Clearly, there are only two TRI momenta in the first BZ that are invariant under the $\mathcal{C}_4$ operation:
$(\pi, 0)$ and $(0, \pi)$. At these two momenta,
all energy states can be labeled by the eigenvalues of $\mathcal{C}_4$
\begin{eqnarray}
\begin{aligned}
\xi_{mn}=e^{i\pi[m+2(1-n)]/4},\quad m,n=\pm1.
\label{ximn}
\end{aligned}
\end{eqnarray}
Physically, $m$ is the eigenvalue of $s_z$ denoting up and down spins of each Kramers doublet,
and $n$ is the eigenvalue of $\tau_z$ denoting particle and hole states in the BdG formalism.
Thus, $\xi_{mn}$ transforms as follows
\begin{eqnarray}
{\Theta}: (m,n)\rightarrow(-m,n),\quad
{\Xi}:(m,n)\rightarrow(-m,-n).
\label{ximn}
\end{eqnarray}
Evidently, states with $\mathcal{C}_4$ eigenvalues $e^{\pm i\pi/4}$ (or $e^{\pm i3\pi/4}$)
form $n=1$ (or $n=-1$) Kramers doublets, and these two groups are related by the particle-hole symmetry.

We now use the eigenvalues of $\mathcal{P}$ to establish the bulk topological invariant,
built upon the algebra in Eq.~(\ref{Comm}).
(i) As $\mathcal{P}$ commutes with both $\Theta$ and $\mathcal{C}_4$,
the two states in each Kramers doublet ($m=\pm1$) must share the same $\mathcal{P}$ eigenvalue.
(ii) In an anomaly-free lattice model, all the particle ($n=1$)  or hole ($n=-1$) states must have an even number
of Kramers doublets with $\mathcal{P}$ eigenvalue $-1$ at the two $\mathcal{C}_4$ invariant momenta.
(iii) As $\mathcal{P}$ commutes with $\Xi$, the particle and hole states of opposite energies
must have the same number of Kramers doublets with $\mathcal{P}$ eigenvalue $-1$
at each $\mathcal{C}_4$ invariant momentum.

Therefore, there is a $\mathbb{Z}_2$ topological invariant for the studied second-order TSC:
the parity of the number of negative-energy Kramers doublets
with $\mathcal{C}_4$ eigenvalues $e^{\pm i\pi/4}$ ({\em i.e.}, the particle states)
and $\mathcal{P}$ eigenvalue $-1$ at the two $\mathcal{C}_4$ invariant momenta.
The odd (or even) parity determines the presence (or absence) of
the four pairs of symmetry-protected Majorana corner states. 
This $\mathbb{Z}_2$ criterion can be applied to any square-shaped case ({\em e.g.}, Fig.~\ref{fig2})
regardless of orientation, as long as it respects the $\mathcal{C}_4$ symmetry.
Since the energy reference is the Fermi energy
and the pairing vanishes at the $\mathcal{C}_4$ invariant momenta,
the $\mathbb{Z}_2$ criterion is reduced to the parity of the number of band inversions
at $(\pi,0)$ and $(0,\pi)$ when the Fermi energy is in the band gap of $h^{\rm TI}$.
This is consistent with our numerical results summarized in Table~\ref{Table}. 
We point out that the $\mathbb{Z}_2$ invariant can alternatively be proved by a bulk-boundary correspondence~\cite{supp}.

\begin{figure}[!t]
\includegraphics[width=0.93\columnwidth]{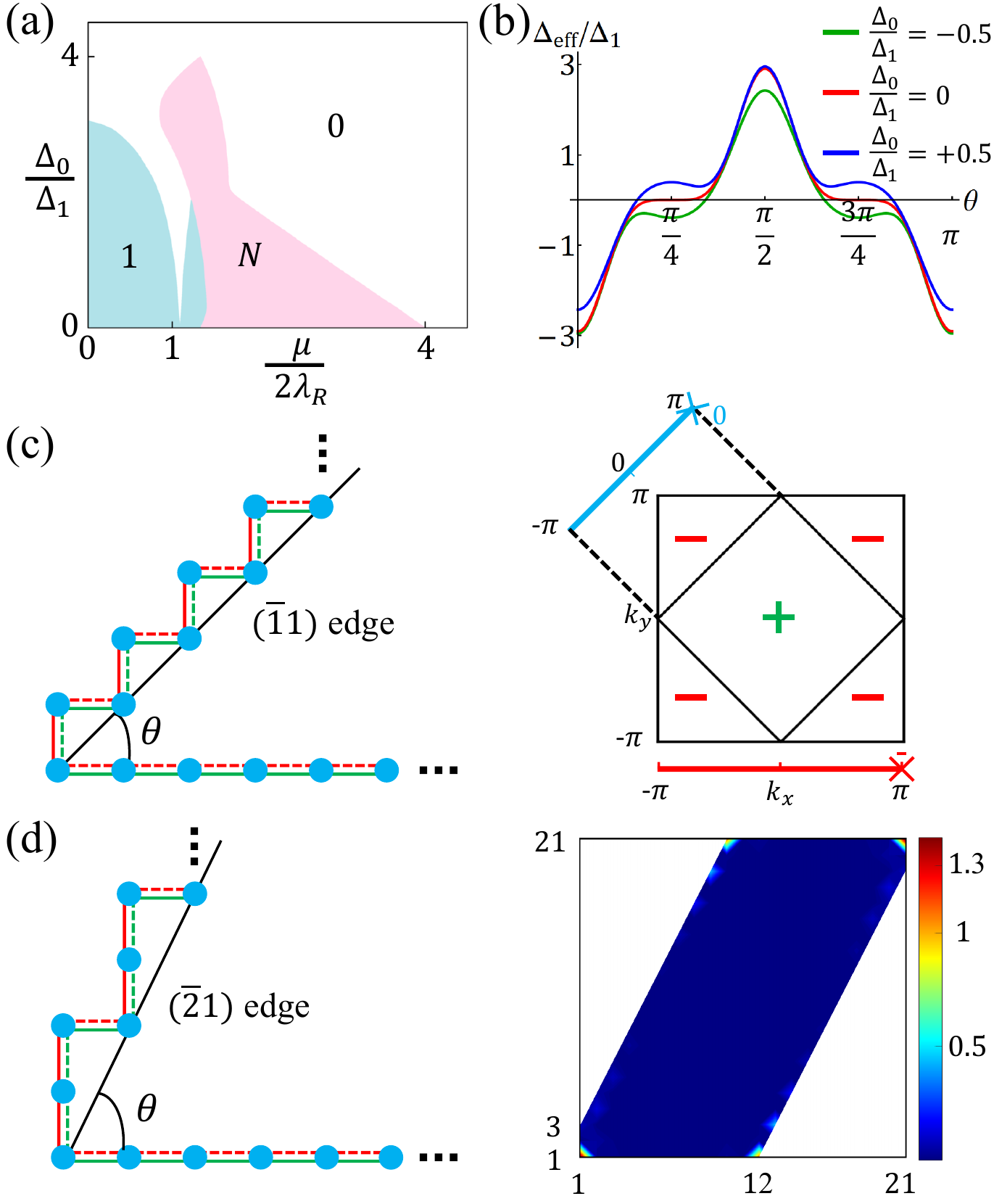}
\caption{(a) Phase diagram of the second-order TSC vs. $\Delta_0$ and $\mu$,
distinguishing the topological ($1$), trivial ($0$), and nodal ($N$) phases.
(b) Effective pairing $\Delta_{\rm eff}$ acquired by the edge state vs. the tilted edge orientation $\theta$.
(c) Schematic of $\Delta_{\rm eff}=0$ for the $({\bar 1}1)$ edge state.
(d) Spatial probability distribution revealing the four pairs of corner Majoranas
for the $11\times21$ parallelogram size TI.
We have chosen $t_1=1$, $t=2$, $\lambda_R=1.5$, $\Delta_1=1$ in (a)-(d),
$\mu=0$ in (b)-(d), and $\Delta_0=0$ in (c)-(d).}
\label{fig3}
\end{figure}

\indent\textcolor{blue}{\em Phase diagram.}---More generally, the protection of
a local Majorana Kramers pair only requires the time-reversal and particle-hole symmetries.
The former dictates the Kramers degeneracy,
whereas the latter pins the Kramers pair to the zero energy.
Consequently, the Majorana Kramers pair at each corner in Fig.~\ref{fig2}f is robust against
the rotational and inversion symmetry breaking, {\em e.g.}, by a nonzero $\Delta_0$ or a tilted edge orientation,
as long as the perturbation does not close the bulk energy gap or hybridize Majoranas at different corners.

Figure~\ref{fig3}a shows the phase diagram of our model~(\ref{Hbdg}) versus $\Delta_0$ and $\mu$
for a case in which the 2D TI has only one band inversion at $(\pi,0)$.
Since the phase diagram is symmetric around $\Delta_0=0$ and $\mu=0$,
we focus on the case with $\Delta_0\geq0$ and $\mu\geq0$.
At $\Delta_0=0$ the $\mathcal{C}_4$ symmetry
is intact, and the established topological criterion is applicable.
When $\mu$ lies in the band gap of the TI, the phase is topologically nontrivial,
as there is one negative-energy Kramers doublet
with $\mathcal{C}_4$ eigenvalues $e^{\pm i\pi/4}$ and $\mathcal{P}$ eigenvalue $-1$
at $(\pi,0)$ whereas none at $(0,\pi)$ --- the total parity is odd.
When $\mu$ crosses the conduction band,
if the Fermi surface and the nodal lines of $\Delta_{\bm k}$ intersect,
the phase becomes nodal. When $\mu$ is beyond the entire conduction band, the phase must be trivial;
at each $\mathcal{C}_4$ invariant momentum the one Kramers doublet
with $\mathcal{C}_4$ eigenvalues $e^{\pm i\pi/4}$ and $\mathcal{P}$ eigenvalue $-1$
is below the Fermi energy --- the total parity is even.

As $\Delta_0$ increases from zero, the topological character remains
unless there is a gap closure in the 2D bulk or at a 1D edge, as shown in Fig.~\ref{fig3}a.
When $\Delta_0$ is above a threshold $\lesssim 4\Delta_1$,
the induced pair potential has a uniform sign at all edges,
and the edge superconductivity must be trivial without the sign reversal~\cite{Qi2,FZ1}.
As $\mu$ increases from zero, the $(0{\bar 1})$-edge-state Fermi points move from $k_x=\pi$ to $k_x=0$,
whereas the $({\bar 1}0)$-edge-state Fermi points move from $k_y=0$ to $k_x=\pi$.
For $\Delta_0=0$, the switch occurs at $k^c=\pi/2$ of both edges,
resulting in an accidental gap closure near $\mu=2\lambda_R\sin k^c$.
For $\Delta_0>0$, however, $k_x^c \neq k_y^c$, and the edge gap closes and reopens twice
near $\mu_{x,y}=2\lambda_R\sin k_{x,y}^c$;
each reverses the pairing sign of one of two edge states.
As a result, the phase is topologically nontrivial before and after the two gap closures but trivial in between.
This explains the presence of two topological domes in Fig.~\ref{fig3}a.
In addition, the larger the value of $\Delta_0$, the smaller the nodal lines of $\Delta_{\bm k}$,
giving rise to the shrinking of the nodal regime in the phase diagram with increasing $\Delta_0$.

Now we reveal the stability of Majorana corner states
against the edge orientations of 2D TI. Without loss of generality, we study the $\mu=0$ case for
the corner connecting the $(0{\bar 1})$ edge and the edge with a tilted angle of $\theta$ from it.
Since the $(0{\bar 1})$ edge state acquires a negative  pairing,
the positive or negative sign of $\Delta_{\rm eff}$, {\em i.e.}, the pairing acquired by the tilted-edge state,  
determines whether the phase is topologically nontrivial or trivial~\cite{Qi2,FZ1}.
Fig.~\ref{fig3}b displays $\Delta_{\rm eff}$ versus $\theta$,
as extracted from our numerical calculations.
For $\Delta_0=0$, the two critical points occur at $\theta=\pi/4$~and~$3\pi/4$
where $\Delta_{\rm eff}$ vanishes, as illustrated in Fig.~\ref{fig3}c.
Fig.~\ref{fig3}d sketches a parallelogram-shaped TI with $\theta=\arctan2$
and features the four pairs of corner Majoranas,
demonstrating that this case is still a second-order TSC.
Generally, for a positive (or negative) $\Delta_0$, the topological regime is enlarged (or suppressed).

\indent{\color{blue}{\em Experiment} \& {\em discussion.}}---Our scheme requires the proximity coupling between
a 2D TI and a nodeless $s_{\pm}$-wave Fe-based superconductor (FSC).
The intrinsic proximity effect between the FSC and its topological surface state
has been experimentally observed in FeTe$_{1-x}$Se$_x$ ($x\!=\!0.45$) with $T_c\!=\!15$~K~\cite{Fe1,Fe2}.
In our extrinsic case, to maintain the phase coherence of ${\bm k}$-dependent pairing,
a lattice match between the TI and FSC is desired.
IV-VI monolayers have been identified as tunable 2D TIs~\cite{Liu,Wrasse,Wan},
with independently controllable band inversions at $(\pi,0)$ and $(0,\pi)$~\cite{Wan}.
The optimized monolayer PbS has a square lattice constant of $4.03$~\AA~\cite{Wan},
comparable to $3.95$--$4.05$~\AA~of iron pnictides~\cite{Dai}.
Thus, the (001) PbS-FSC heterostructure can be a candidate system to explore the proposed second-order TSC.

As critical advantages, our scheme requires neither a $\pi$ Josephson junction nor a magnetic field,
and the FSC can enjoy a critical temperature as high as $56$~K in Sr$_{0.5}$Sm$_{0.5}$FeAsF~\cite{XHC}. 
In probing the Majoranas, because of the Kramers degeneracy, the zero-bias tunneling conductance
is anticipated to be $4e^2/h$~\cite{FZ1} at each corner when the TSC is grounded.
In the future, it would be interesting to braid the second-order Majoranas 
that are anticipated to have non-abelian statistics~\cite{NA1,NA2} 
and to discover other higher-order anyons~\cite{JA1,JA2}.

Although we focus on the scenario in which the 2D TI has one band inversion at $(\pi,0)$,
Majorana corner states can exist in other scenarios.
(i) Consider a TI with one band inversion at $(0,0)$ or $(\pi,\pi)$.
For a $d_{x^2-y^2}$-wave pairing (if nodeless), very similar physics is anticipated; 
for an $s_{\pm}$-wave pairing, 
corner Majoranas are possible only if the structure strongly breaks the rotational symmetry.
(ii) Consider a TI with two band inversions, {\em e.g.}, at $(0,0)$ and $(\pi,0)$. For either aforementioned pairing, 
both the $(01)$ and $(0{\bar 1})$ edges have dual helical edge states at $k_x=0,\pi$ and may become TSCs 
while both the $(10)$ and $({\bar 1}0)$ edges are trivial.
Interesting, monolayer WTe$_2$ as a TI up to $100$~K~\cite{qsh1,qsh2}
has one band inversion at $(0,0)$ and may be exploited for scenario (i).

\indent{\color{blue}\em Acknowledgments.}---This work was supported by UT-Dallas research enhancement funds (QW and FZ)
and NSF under award number DMR-1653769 (YML).

\indent{\color{blue}\em Note added.}---During the finalization we became aware of two complementary
studies: one on edge theories~\cite{new1} and the other on different pairing symmetries~\cite{new2}.

\end{document}


\title{
       Supplementary Material for ``High-Temperature Majorana Corner States''
       }
 \author{Qiyue Wang}
\affiliation{Department of Physics, University of Texas at Dallas, Richardson, Texas 75080, USA}
\author{Cheng-Cheng Liu}
\affiliation{School of Physics, Beijing Institute of Technology, Beijing 100081, China}
\author{Yuan-Ming Lu}
\affiliation{Department of Physics, Ohio State University, Columbus, Ohio 43210, USA}
\author{Fan Zhang}
\affiliation{Department of Physics, University of Texas at Dallas, Richardson, Texas 75080, USA}
\maketitle

\subsection{An alternative proof of bulk topological invariant}

Here we prove the bulk $\mathbb{Z}_2$ topological invariant for the second-order TSCs 
using the bulk-boundary correspondence. Specifically, we consider how the eight Majorana corner states 
of a square-shaped open system are coupled together to form a $\mathcal{C}_4$ invariant 
closed system described by the bulk band structure. 
This allows us to establish a one-to-one correspondence between the eight Majorana corner states and 
the $\mathcal{P}$ eigenvalues of bulk bands at the $\mathcal{C}_4$ invariant momenta.

First consider a $\mathcal{C}_4$ symmetric open system on a $N\times N$ square lattice, 
For a second-order TSC, by definition, all bulk and edge states are fully gapped 
except for one zero-energy Majorana Kramers pair at each of the four corners. 
We thus focus on the eight zero-energy levels in the open system.
Without loss of generality, the symmetry implementations on these eight levels can be written as 
\begin{eqnarray}
\Theta&=&i s_y\cdot\mathcal{K},\notag\\
\Xi&=&s_0\otimes\left( \begin{array}{cccc} 1&&&\\&-1&&\\&&1&\\&&&-1\end{array}\right)\cdot\mathcal{K},\notag\\
\mathcal{P}&=&i s_y\otimes\left( \begin{array}{cccc}&&-1&\\&&&-1\\1&&&\\&1&&\end{array}\right),\notag\\
\mathcal{C}_4&=&s_0\otimes\left( \begin{array}{cccc}&&&-1\\1&&&\\&1&&\\&&1&\end{array}\right),\notag
\end{eqnarray}
where $\bm s$ are the Pauli matrices for the spin space (of each Kramers doublet). 
They are determined by the algebra in Eq.~(5) and completely fixed up to a local unitary transformation.

Now we glue the open boundaries to form a closed system while preserving all the symmetries. 
The eight Majorana zero modes must be gapped out by a mass term that respects all the symmetries. 
It is straightforward to show that the only mass term is $\mathcal{P}$, {\em i.e.}, the inversion symmetry operator. 

Note that there is a subtlety in choosing the boundary conditions 
for a $\mathcal{C}_4$ symmetric periodic system. By choosing $N$ to be an odd integer, 
if we label the coordinates of the unit cell at the center of square as $(x,y)=(0,0)$, 
the four corners are then located at $\frac{N-1}{2}(\pm1,\pm1)$. 
Due to the $(-1)^{x+y}$ factor in the bulk $\mathcal{C}_4$ operation, 
the $\mathcal{C}_4$ symmetry dictates the boundary conditions along the $x$- and $y$-directions must be reversed: 
it is either a periodic boundary condition along the $x$-direction (PBC$_x$) 
and an anti-periodic boundary condition along the $y$-direction (APBC$_y$), 
or the other way around, {\em i.e.}, APBC$_x$ and PBC$_y$. 
On an odd by odd square lattice, due to quantization of momentum under the proper boundary conditions, 
only one of the four TRI momenta will be occupied, 
and it can be either of the two $\mathcal{C}_4$ invariant momenta: 
\begin{eqnarray}
\{\mbox{PBC}_x,~\mbox{APBC}_y\}&\rightarrow&(0,\pi),\notag\\
\{\mbox{APBC}_x,~\mbox{PBC}_y\}&\rightarrow&(\pi,0).\notag
\end{eqnarray}
The occupation of all other momenta remains the same between these two different boundary conditions.

Thus, the choice of boundary conditions will pick one of the two $\mathcal{C}_4$ invariant momenta. 
To switch from \{PBC$_x$,~APBC$_y$\} to \{APBC$_x$,~PBC$_y$\}, the energy gap must close, 
corresponding to the sign change of the mass term. 
Now that the mass term has been proved to be the inversion symmetry operator,  
this also reverses the bulk topological invariant defined previously.
Therefore, we have shown that for our gapped superconductors with protected Majorana corner states, 
the bulk invariant must be nontrivial. This establishes our proof of the bulk topological invariant.